# Blind Monaural Source Separation on Heart and Lung Sounds Based on Periodic-Coded Deep Autoencoder

Kun-Hsi Tsai[1], Wei-Chien Wang[1], Chui-Hsuan Cheng[1], Chan-Yen Tsai[1], Jou-Kou Wang[2], Tzu-Hao Lin[3],
Shih-Hau Fang[4], Li-Chin Chen[5], Yu Tsao[5]

*Abstract*—Auscultation is the most efficient way to diagnose cardiovascular and respiratory diseases. To reach accurate diagnoses, a device must be able to recognize heart and lung sounds from various clinical situations. However, the recorded chest sounds are mixed by heart and lung sounds. Thus, effectively separating these two sounds is critical in the pre-processing stage. Recent advances in machine learning have progressed on monaural source separations, but most of the well-known techniques require paired mixed sounds and individual pure sounds for model training. As the preparation of pure heart and lung sounds is difficult, special designs must be considered to derive effective heart and lung sound separation techniques. In this study, we proposed a novel periodicity-coded deep auto-encoder (PC-DAE) approach to separate mixed heart-lung sounds in an unsupervised manner via the assumption of different periodicities between heart rate and respiration rate. The PC-DAE benefits from deep-learning-based models by extracting representative features and considers the periodicity of heart and lung sounds to carry out the separation. We evaluated PC-DAE on two datasets. The first one includes sounds from the Student Auscultation Manikin (SAM), and the second is prepared by recording chest sounds in real-world conditions. Experimental results indicate that PC-DAE outperforms several well-known separation works in terms of standardized evaluation metrics. Moreover, waveforms and spectrograms demonstrate the effectiveness of PC-DAE compared to existing approaches. It is also confirmed that by using the proposed PC-DAE as a pre-processing stage, the heart sound recognition accuracies can be notably boosted. The experimental results confirmed the effectiveness of PC-DAE and its potential to be used in clinical applications.

K.-H. Tsai, W.-C. Wang, C.-H. Cheng, C.-Y. Tsai are with the Medical Department, Imediplus Inc. Taipei, Taiwan ({peter.tsai; ethan.wang;hammer.cheng;mike.tsai}@imediplus.com).
L.-C. Chen and Y. Tsao are with the Research Center for Information Technology Innovation, Academia Sinica, Taipei 11529, Taiwan. ({li.chin; yu.tsao)@citi.sinica.edu.tw).
J.-K. Wang is with Department of Pediatrics, National Taiwan University Hospital, Taipei, Taiwan (jkww@ntuh.gov.tw).
S.-H. Fang is with Department of Electric Engineering, Yuan Ze University, Taoyuan, Taiwan (shfang@saturn.yzu.edu.tw)
T.-H. Lin is with Biodiversity Research Center, Academia Sinica, Taiwan (schonkopf@gmail.com)

## I. INTRODUCTION

Recently, biological acoustic signals have been enabling various intelligent medical applications. For example, the biological acoustic signals of the heart and lung can facilitate tasks such as diagnosing the cardiovascular and respiratory diseases, and monitoring the sleep apnea syndrome [1-8]. Previous studies have already investigated the physical models of the heart and lung sound generation and classification mechanisms. For example, signal processing approaches (e.g., normalized average Shannon energy [9] and high-frequency-based methods [10]) and machine-learning-based models (e.g., neural network (NN) classifiers [11] and decision trees [12]) have been used to perform heart disease classification based on acoustic signals. In addition, the information of S1–S2 and S2–S1 intervals has been adopted to further improve the classification accuracies [12], [13]. On the other hand, Gaussian mixture model [13] NN classifiers [14], and support vector machines[15] along with various types of acoustic features (e.g., power spectral density values, Hilbert-Huang transform[16]) have been utilized to carry out lung sound recognition [17, 18]. However, medical applications using such biological acoustic signals still face several challenges.

To reach accurate recognition, sound separation is one of the most important pre-processing. Because the measured signal is usually a mixed version of the heart and lung sounds, and pure heart/lung acoustic signals is generally not accessible, effectively separating heart and lung sounds is very challenging. The frequency ranges of normal heart sounds (first(S1) and second(S2) heart sound) is mainly 20-150 Hz, and some high-frequency murmurs may reach to 100-600 Hz, or even to 1000 Hz [19]. On the other hand, the frequency range of normal lung sounds is 100-1000 Hz (tracheal sounds range from 850 Hz to 1000 Hz), abnormal lung sound as adventitious sounds of wheeze span a wide range of frequencies variation of 400-1600 Hz, and the range for crackle and rales is 100-500 Hz [20, 21]. Therefore, the frequency range of the heart and lung sounds can be highly overlapped. This results in interference between the acoustic signals and may degrade the auscultation and monitoring performance. With an increasing demand for various acoustic-signal-based medical applications, effective heart and lung sound separation techniques have become fundamental, although challenging.

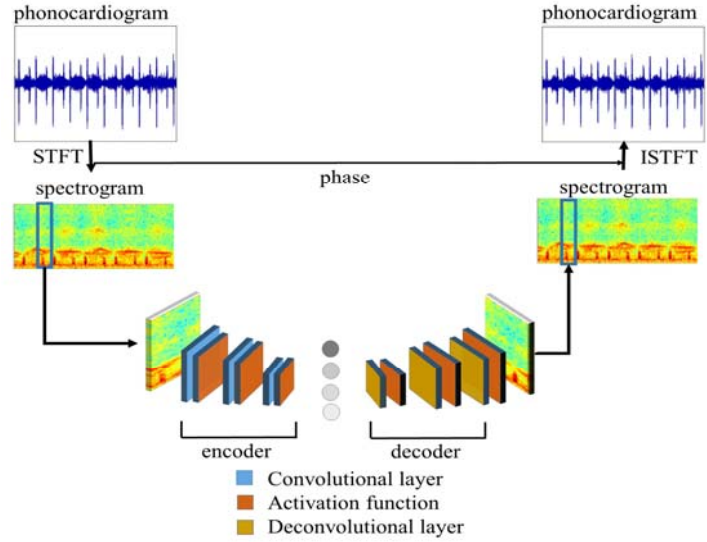

Fig. 1. The convolutional deep autoencoder (DAE(C)) architecture.

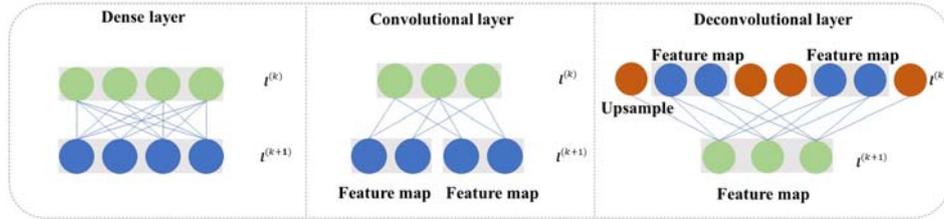

Fig. 2. Relation between hidden layers in a fully connected layer, convolutional layer, and deconvolutional layer.

Sound separation techniques for heart and lung have been studied extensively, and numerous methods have been proposed so far. For example, the study [22-26] focuses on the adaptive filtering approach while Mondal et al. [27, 28] use the empirical mode decomposition methods. Hossain and Hadjileontiadis et al. [29, 30] proposed to use the discrete wavelets transform approach to filter interference. Pourazad et al. [31] derived an algorithm that transforms the signal to time-frequency domain (STFT) and combined with the continuous wavelets transform (CWT) to filter out heart sound components by a band-pass filter.

However, the above-mentioned traditional filtering approaches encounter difficulties due to the overlapped frequency bands. The works in [32-34] proposed the blind source separation algorithms, including independent component analysis (ICA) and its extensions, in which the prior knowledge of sources is not required. Nevertheless, the ICA-based methods require at least two sensors and thus, do not work for the devices having only single-channel [35-37]. The assumption of independence between heart sound sources is somehow optimistic.

Recently, the supervised monaural (single-channel) nonnegative matrix factorization (NMF) was adopted to separate different sources [35, 38]. It was recognized for its capability of handling overlapping frequency bands [39, 40]. More recently, deep learning approaches have been used for source separation [40-43]. Although these deep models directly dismantle the mixture source into the target ones and outperform the NMF approach, those frameworks were subject to supervised training data. However, in biomedical applications, the training data of pure heart/lung acoustic signals is difficult or too expensive to measure.

To overcome the mentioned challenges, this paper proposes a periodicity-coded deep autoencoder (PC-DAE) approach, an unsupervised-learning-based mechanism to effectively separate the sounds of heart and lung sources. The proposed algorithm first adopts the DAE model [40, 44-46] to extract highly expressive representations of the mixed sounds. Next, by applying the modulation frequency analysis (MFA) [47] on the latent representations, we can group the neurons based on their properties in the modulation domain and then perform separation on the mixed sound. The advantage of PC-DAE is that the labeled training data (more specifically, paired mixed sounds and individual pure sounds) are not required as compared to the typical learning-based approaches. It benefits from the periodicity structure to provide superior separation performance than the traditional methods.

The remainder of this paper is organized as follows. In Section 2, we will review the NMF and DAE algorithms. In Section 3, the proposed PC-DAE will be introduced in detail. In Section 4, we present the experimental setup and results, where two datasets were designed and used to test the proposed PC-DAE model. The first one is phonocardiogram signals from the Student Auscultation Manikin (SAM) database [48], and the second one is prepared in a real-world condition. Experimental results confirm the effectiveness of PC-DAE to separate the mixed heart-lung sounds with outperforming related works, including direct-clustering NMF (DC-NMF) [35], PC-NMF [49], and deep clustering (DC) [45], in terms of three standardized evaluation metrics, qualitative comparisons based on separated waveforms and spectrograms, and heart sound recognition accuracy.

## II. RELATED WORKS

Numerous methods have been proposed to separate the heart and lung sound signals. Among them, the NMF is a notable one that has been applied to separate different sounds [35, 38]. The DAE model is another well-known approach. Based on the model architecture, the DAE can be constructed by a fully connected architecture, termed DAE(F), or by a fully convolutional architecture, termed DAE(C). In this section, we provide a review of the NMF algorithm, DAE(F), and DAE(C) models.

### A. Non-negative matrix factorization (NMF)

The conventional NMF algorithm factorizes the matrix $V$ into two matrices, a dictionary matrix $W$ and an encoded matrix $H$. The product of the $W$ and $H$ approximates matrix $V$. All the matrices entries are nonnegative. The NMF-based source separation can be divided into two categories, namely supervised (where individual source sounds are provided) and unsupervised (where individual source sounds are not accessible). For supervised NMF-based approaches, a pre-trained, fixed spectral matrix $W^S$, where $W^S = [W_1^S ... W_A^S]$, and $A$ is the number of sources, which consists of the characters of each sound source is previously required [35, 50]. To process NMF, first, the recording that consists of multiple sounds was factorized by NMF into $W^S$ and $H^T$. Then $H^T$ is divided into $A$ blocks: $H^T = [H_1^T ... H_A^T]$. Through multiplying $W_i^S$ and $H_i^T$ ($i=1,...A$), we obtain individual sound sources.

For unsupervised NMF-based approaches, since individual source sounds are not available, some statistical assumptions must apply. An intuitive approach is to cluster the vectors in $H$ to several distinct groups. A particular sound can be reconstructed by a group of vectors in $H$ along with $W$. The work of Lin et al [49], on the other hand, designed PC-NMF using another concept, which is to incorporate the periodicity property of distinct source sounds into the separation framework. More specifically, PC-NMF considers the encoded matrix $H^T$ as the time vectors and uses the nature of periodical differences to separate the biological sounds. Because heart sound and lung sounds are different in periodic characters (heart rate and respiration rate are very different), the mixed heart-lung sound is separated through a PC-NMF model, as will be presented in Section 4.

### B. Deep Autoencoder (DAE)

The DAE has two components, an encoder $E(\cdot)$ and a decoder $D(\cdot)$. Figure 1 shows the architecture of a DAE(C) model. Consider the encoder and decoder to have $K_E$ and $K_D$ layers, respectively, the total number of layers in the DAE is $K_{All} = K_E + K_D$. The encoder encodes the input $x$ to the middle latent space $l^{(K_E)}$ ($l^{(K_E)} = E(x)$), and the decoder reconstructs the input by ($\hat{x} = D(l^{(K_E)})$). The reconstructed output $\hat{x}$ is expected to be approximately equal to $x$. The mean squared error (MSE) is generally used to measure the difference between $\hat{x}$ and $x$. Minimizing the value of MSE is the goal to train the DAE model. As mentioned earlier, by using fully connected and fully convolutional architectures, we can build DAE(F) and DAE(C), respectively [51-53]. Fig. 2 shows the neuron connections of the $k$-th and ($k$+1)-th layers for the two types of DAE. Fig. 2(a) presents the fully-connected layer, where each neuron in the ($k$+1)-th layer is fully-connected with all neurons in the $k$-th layer. Fig. 2 (b) and (c), respectively, present the convolutional and deconvolutional connections, where each neuron in the ($k$+1)-th layer is partially-connected with the neurons in the $k$-th layer. As can be seen from Fig. 2(a), the DAE(F) forms the encoder and decoder by fully-connected units, which is shown in Eqs. (1) and (2), $W_E^{(k)}$ and $W_D^{(k)}$ represent the encoding and decoding matrix, $b_E^{(k)}$ and $b_D^{(k)}$ are the bias terms:

$$l^{(1)} = \sigma(W_E^{(0)} x + b_E^{(0)})$$
$$l^{(k+1)} = \sigma(W_E^{(k)} l^{(k)} + b_E^{(k)}) \quad k = 1,..., K_E\text{-}1, \quad (1)$$

where $l^{(k)} \in R^{M \times 1}$, and $M$ stands for the total number of neurons in the latent space. For the decoder, we have

$$l^{(k+1)} = \sigma(W_D^{(k)} l^{(k)} + b_D^{(k)}), k = K_D ..., (K_{All}\text{-}1)$$
$$\hat{x} = \sigma\left(W_D^{(K_{All})} l^{(K_{All})} + b_D^{(K_{All})}\right). \quad (2)$$

In DAE(C), the encoder is formed by convolutional units, as shown in Eq. (3), that executes the convolutional function $F_{Conv}(\cdot)$. Each encoded layer has $J$ filters: $\{W_1, ..., W_J\}$; $W_j \in R^{L \times 1}$, $L$ is the kernel size, and $W_{ji}$ is the $i$-th channel of $W_j$, where $W_{ji} = (w_1, ..., w_I)$. Each neuron in the ($k$+1)-th layer's feature map, $l_j^{(k+1)}$, is the summation of the element-wised product of $W_j$ and receptive field of all previous feature maps $l^{(k)}$ by convolution operation, and $b_j^{(k)}$ denotes the bias term. The corresponding convolution operation is shown in Fig. 3 (a). The decoder is formed by a deconvolutional unit, as shown in Eq. (4). During deconvolution, all of the $k$-th layer's feature maps $l^{(k)}$ first go through the zero-padding and then deconvolution processes (with function $F_{Deconv}(\cdot)$). Each decoded layer has $J$ filters: $\{W_1, ..., W_J\}$; $W_j \in R^{L \times 1}$, $L$ is the kernel size, and $W_{ji}$ is the $i$-th channel of $W_j$, where $W_{ji} = (w_1, ..., w_I)$. Each neuron in the ($k$+1)-th layer, $l_j^{(k+1)}$, is the summation of the element-wised product of $W_j$ and receptive field of all previous feature maps $l^{(k)}$ by deconvolution operation, and $b_j^{(k)}$ denotes the bias terms. The corresponding deconvolution operation is shown in Fig. 3 (b).

$$l_j^{(1)} = \sigma\left(F_{Conv}(W_{ji}^{(0)}, x) + b_{E_j}^{(0)}\right)$$
$$l_j^{(k+1)} = \sigma\left(\sum_{i=1}^{I} F_{Conv}(W_{ji}^{(k)}, l^{(k)}) + b_{E_j}^{(k)}\right) \quad (3)$$

where $l_j^{(k)}$ is the $j$-th feature map in the $k$-th layer, and $I$ is the total number of channels. For the decoder, we have

$$l_j^{(k+1)} = \sigma\left(\sum_{i=1}^{I} F_{Deconv}(W_{ji}^{(k)}, l^{(k)}) + b_j^{(k)}\right)$$
$$\hat{x} = \sigma\left(F_{Deconv}(W_{ji}^{(K_{All})}, l^{(K_{All})}) + b_j^{(K_{All})}\right) \quad (4)$$

where $K_{All}$ denotes the total number of layers in the DAE(C).

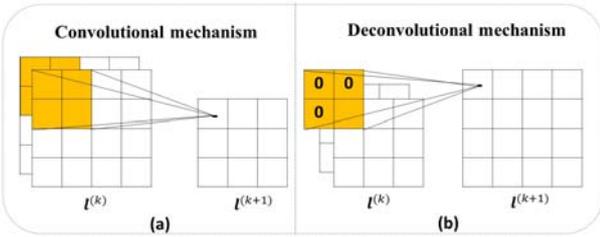

Fig. 3 (a) Convolutional and (b) deconvolutional operations.

## III. THE PROPOSED METHOD

The proposed PC-DAE is a DAE-based unsupervised sound source separation method. When performing separation, the recorded sounds are first transformed to spectral-domain and phase parts via short-time Fourier transform (STFT). The spectral features are converted to log power spectrum (LPS) [52], where $X = [x_1, \ldots, x_n, \ldots, x_N]$ denotes the input, and $N$ is the number of frames of $X$. Then the DAE encodes the mixed heart-lung LPS by $E(\cdot)$ to convert $X$ to the matrix of latent representations, $L^{(K_E)} = [l_1^{(K_E)}, \ldots, l_n^{(K_E)}, \ldots, l_N^{(K_E)}]$. The decoder, $D(\cdot)$, then reconstructs the latent representations back to original spectral features. The back-propagation algorithm [54] is adopted to train the DAE parameters to minimize the MSE scores. Because the input and output are the same, the DAE can be trained in an unsupervised manner.

With the trained DAE, the periodic analysis is applied to the latent representations to identify two disjoint portions of neurons corresponding to heart and lung sounds. The basic concept is to consider the temporal information of different periodic sources. Moreover, to classify the temporal information by periodicity, the coded matrix is transformed into periodic coded matrix $P$ via modulation frequency analyzer (MFA). Here, we adopted the discrete Fourier transform (DFT) to perform MFA. The periodic coded matrix presents clear periodicity characteristics. Because heart sound and lung sound have different periodicity, the coded matrix can be separated to heart coded matrix and lung coded matrix from the whole encoded matrix, $P$. Afterwards, each source coded matrix is transformed by the decoder and reconstructed to obtain the LPS sequences of the separated heart sound $Y^{heart}$ and lung sound $Y^{lung}$. The output LPS features are then converted back to waveform-domain signals by applying inverse short-time Fourier transform (ISTFT).

### A. Periodic Analysis Algorithm

In this section, we present the details of the MFA. Fig. 4 illustrates the overall PC-DAE framework. First, we train a DAE(F) or DAE(C) model with the encoder and decoder as shown in Eqs. (1) and (2) or Eqs. (3) and (4), respectively. Then, we input the sequence of mixed heart-lung sounds, $X$, to obtain the latent representations. The collection of latent representations and the time sequence are the matrix $L = \{l_1^{(K_E)}, l_2^{(K_E)}, \ldots l_N^{(K_E)}\}$. Thus, we obtain

$$L = [E(x_1) \quad \ldots \quad E(x_n) \quad \ldots \quad E(x_N)]$$
$$= \begin{bmatrix} \begin{bmatrix} l_{11}^{(K_E)} \\ \vdots \\ l_{j1}^{(K_E)} \\ \vdots \\ l_{M1}^{(K_E)} \end{bmatrix} \ldots \begin{bmatrix} l_{1n}^{(K_E)} \\ \vdots \\ l_{jn}^{(K_E)} \\ \vdots \\ l_{Mn}^{(K_E)} \end{bmatrix} \ldots \begin{bmatrix} l_{1N}^{(K_E)} \\ \vdots \\ l_{jN}^{(K_E)} \\ \vdots \\ l_{MN}^{(K_E)} \end{bmatrix} \end{bmatrix}, \quad (5)$$

where $L \in R^{M \times N}$, $j$ is the neuron index, where $1 \leq j \leq M$, and $n$ is the time stamp, where $1 \leq n \leq N$, and $N$ is the total number of frames.

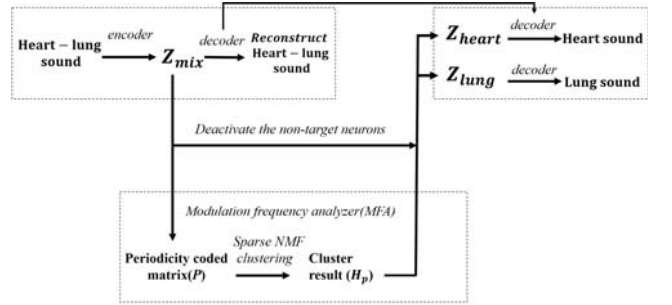

Fig. 4. The PC-DAE Framework.

We assume that among the latent representations, some neurons are activated by heart sound and the others activated by lung sounds. Based on this assumption, we can separate mixed heart-lung sounds in the latent representation space. To determine whether each neuron is activated either by heart or lung sound, we transpose the original $L$ to obtain $Z^{mix} = L^T$ ($T$ denotes matrix transpose). Thus, we obtain

$$Z^{mix} = [z_1^{mix}, \ldots, z_j^{mix}, \ldots, z_M^{mix}],$$
where
$$z_j^{mix} = [l_{j1}^{(K_E)}, \ldots, l_{jn}^{(K_E)}, \ldots, l_{jN}^{(K_E)}]^T \quad (6)$$

With $Z^{mix}$, we intend to cluster the entire set of neurons into two groups, one group corresponding to heart sounds and the other to lung sounds. More specifically, when pure heart sound is inputted to the DAE, only one group of neurons corresponding to the heart sounds is activated, and the other group corresponding to the lung sounds is deactivated. When the pure lung sound is inputted to the DAE, on the other hand, the group of neurons corresponding to the lung sounds is activated, and the other group corresponding to the heart sounds is deactivated. The strategy to determine these two groups of neurons is based on the periodicity of heart and lung sounds.

Algorithm 1 shows the detailed procedure of periodic analysis. To analyze the periodicity of each submatrix $z_j^{mix}$, we form the periodic coded matrix $P = [p_1, ..., p_j, ..., p_M]$ by applying the MFA on $z_j^{mix}$, as shown in Eq. (7).

$$p_j = |\text{MFA}(z_j^{mix})|. \qquad (7)$$

When we used DFT to carry out MFA, we have $p_j \in R^{(N/2+1)}$, and $P$ can be clustered into two groups. There are numerous clustering approaches available, and we used the sparse NMF clustering method to cluster the vectors in $P$ into two groups [55]. Eq. (8) shows the clustering process by NMF, which is also achieved by minimizing the error function. On the basis of the largest score in the encoding matrix, $H_p$, of the transposed $P$, the clustering assignment of $Z^{mix}$ can be determined.

$$H_p = \arg\min[\| P - W_p H_p \|^2 + \lambda \| H_p \|], \qquad (8)$$

where $W_p$ represents the cluster centroids, $H_p = [h_1, ..., h_j, ..., h_M]$ represents the cluster membership, $h_j \in R^{k \times 1}$, $k$ is set as the cluster amount of the basis, $\lambda$ represents the sparsity penalty factor, $\|\cdot\|$ represents the L1-norm, and $\|\cdot\|_F^2$ represents the Frobenius distance.

---
**Algorithm 1** : MFA on coded matrix

**Input**: mixed heart-lung coded matrix $Z^{mix}$, where $Z^{mix} \in R^{N \times M}$

**Output**: heart coded matrix $Z^{heart}$, lung coded matrix $Z^{lung}$

1: **for** $j = 1$ to $M$ **do**
2: $\quad p_j = |\text{MFA}(z_j^{mix})|$
3: **end for**

4: Perform clustering on vectors $[p_1, ..., p_M]$ in $P$
5: Obtain labels of $P$: $c = [c_1, ..., c_M]$, where there are only two labels of $c_j$ {heart or lung}.

6: Set $\varepsilon_{min} \in R^{N \times 1}$, where $\varepsilon_{min}$ is a vector whose coefficients are the latent neuron's minimum values
7: **foreach** $t$ = [heart; lung] **do**
8: $\quad$ Initialize $Z^t = Z^{mix}$
9: $\quad$ **for** $j$ (1 to $M$) **do**
10: $\quad\quad$ **if** $c_j \neq t$ **then**
11: $\quad\quad\quad$ do $z_j^t = \varepsilon_{min}$
12: $\quad\quad$ **end if**
13: $\quad$ **end for**
14: $\quad$ **return** $Z^t$
15: **end foreach**

---

On the basis of the $h_j$ of encoding matrix $H_p$, the clustering results $c = [c_1, ..., c_j, ..., c_M]$ is determined by the largest score of $h_j$. In this case, $c_j \in \{\text{heart, lung}\}$, and the cluster results assign to $z_j^{mix}$. According to the assigned clustering result, $Z^{mix}$ is separated to $Z^{heart}$ and $Z^{lung}$ by deactivating the submatrices which do not belong to the target, respectively.

After obtaining the coded matrix of each source, we decode it as Eqs. (9) and (10).

$$Y^{heart} = D(Z^{heart}) \qquad (9)$$
$$Y^{lung} = D(Z^{lung}). \qquad (10)$$

In the proposed approach, we compute the ratio mask of these two sounds, which are defined as Eqs. (11) and (12).

$$M^{heart} = \left(\frac{D(Z^{heart})}{D(Z^{heart}) + D(Z^{lung})}\right) \qquad (11)$$

$$M^{lung} = \left(\frac{D(Z^{lung})}{D(Z^{heart}) + D(Z^{lung})}\right). \qquad (12)$$

With the estimated $M^{heart}$ and $M^{lung}$, we obtain the heart LPS $\hat{Y}^{heart}$ and lung LPS $\hat{Y}^{lung}$ by Eqs. (13) and (14).

$$\hat{Y}^{heart} = M^{heart} \odot D(Z^{mix}) \qquad (13)$$
$$\hat{Y}^{lung} = M^{lung} \odot D(Z^{mix}), \qquad (14)$$

where $\odot$ denotes the element-wise multiplication. Then $\hat{Y}^{heart}$ and $\hat{Y}^{lung}$ along with the original phase are used to obtain the separated heart and lung waveforms.

## IV. EXPERIMENTS

### A. Experimental setups

In addition to the proposed PC-DAE(F) and PC-DAE(C), we tested some well-known approaches for comparison, including direct-clustering NMF (DC-NMF), PC-NMF, and deep clustering based on DAE (DC-DAE). The PC-NMF and PC-DAE shared a similar functionality where the PC-DAE performs clustering on the latent representations for heart and lung sound separation. For a fair comparison, the DC-NMF, PC-NMF, and DC-DAE implemented in this study are carried out in an unsupervised manner. For all the methods, the mixed spectrograms were used as the input, and the separated heart and lung sounds were generated at the output.

The DAE(F) model consisted of seven hidden layers, and the neurons in these layers were 1024, 512, 256, 128, 256, 512, and 1024. The encoder of the DAE(C) model consisted of three convolutional layers. The first layer had 32 filters with a kernel size of 1×4, the second layer had 16 filters with a kernel size of 1×3, and the third layer had 8 filters with a kernel size of 1×3 of the encoder. The decoder comprised of four layers. The first layer had 8 deconvolutional filters with a kernel size of 1×3, the second layer had 16 deconvolutional filters with the kernel size of 1×3, the third layer had 32 deconvolutional filters with a kernel size of 1×4, and the fourth layer had 1 deconvolutional filter with kernel size of 1 × 1. Both convolution and deconvolution units adopt a stride of 1. The rectified linear unit were used in encoder and decoder, and the optimizer was Adam. The unsupervised NMF-based methods were used as the baseline, where the basis number of NMF was set to 20, and the L2 norm was used as the cost function. The NMF approach first decomposes the input spectrogram $V$ into the basis matrix $W$ and the weight matrix $H$, where $W$ serves as the sound basis (including both heart and lung sounds), and $H$ are the weighting coefficients:

$$V_{ij} \approx (WH)_{ij} = \sum_{a=1}^{A} W_{ia}H_{aj}, \quad (15)$$

where $V_{ij}$ is the $ij$-th component of $V$ (a matrix that contains multiple sound sources) and $W_{ia}$ and $H_{aj}$ are the $ia$-th component of $W$ and the $ai$-th component of $H$, respectively.

For unsupervised source separation, the weighting coefficient matrix $H$ is clustered into several distinct groups. When performing separation, the target source of interest can be reconstructed by using the group of vectors in $H$ that corresponds to the target source. Because the clustering is directly applied to the weighting matrix, we refer to this approach as DC-NMF as the first baseline system. Rather than directly clustering, the PC-NMF [49] clusters the vectors in $H$ based on the periodicity of individual sound sources; the PC-NMF was also implemented as the second baseline.

Recently, a deep clustering technique [56] that combines a deep learning algorithm and a clustering process has been proposed and confirmed effective for speech [45] and music [46] separation. The fundamental theory of deep clustering is similar to DC-NMF as the clustering is applied in the latent representations instead of the weighting matrix. Because the deep-learning models first transform the input spectrograms into more representative latent features, the clustering of latent features can provide superior separation results. In this study, we implement a deep clustering approach as another comparative method. We used the model architecture of DAE(C) as the deep-learning-based model when implementing the deep clustering approach; hence, the approach is terms DC-DAE(C).

For all the separation methods conducted in this study, we can obtain separated heart and lung sounds. We used the pure heart and lung sounds as a reference to compute the separation performance and adopted three standardized evaluation metrics, namely signal distortion ratio (SDR), signal to interferences ratio (SIR), and signal to artifacts ratio (SAR) [57] to evaluate the separation performances. In a source separation task, there are three types of noise: (1) noise due to missed separation ($e_{interf}$); noise due to the reconstruction process ($e_{artif}$), and the perturbation noise ($e_{noise}$). The computations of SDR, SIR, and SAR are presented in Eqs. (16)-(19), where $\hat{s}(t)$ is the estimated result and $s_{target}(t)$ is the target.

$$\hat{s}(t) = s_{target}(t) + e_{interf} + e_{noise} + e_{artif} \quad (16)$$

$$\text{SDR} := 10 \, log_{10} \frac{\|s_{target}(t)\|^2}{\|e_{interf} + e_{noise} + e_{artif}\|^2} \quad (17)$$

$$\text{SIR} := 10 \, log_{10} \frac{\|s_{target}(t)\|^2}{\|e_{interf}\|^2} \quad (18)$$

$$\text{SAR} := 10 \, log_{10} \frac{\|s_{target}(t) + e_{interf} + e_{noise}\|^2}{\|e_{artif}\|^2}. \quad (19)$$

For all of these three metrics, higher scores indicate better source separation results.

We conducted experiments using two datasets. In the first dataset, the heart and lung sounds were collected by SAM, which is a standard equipment in teaching and learning heart and lung sounds [48]. Fig. 5 shows the model of SAM. The SAM attempts to simulate the real human body and has many speakers inside its body corresponding to organ's positions. The SAM can generate clean heart sound or lung sound in different locations. We used the iMEDIPLUS electronic stethoscope [58] to record heart and lung sounds in an anechoic chamber. The heart sounds used in this experiment included normal heart sounds with two beats (S1 and S2). The lung sounds in this experiment included normal, wheezing, rhonchi, and stridor sounds. Both heart and lung sounds were sampled at 8k Hz. The two sounds were mixed at different signal to noise ratio (SNR) levels (-6 dB, -2 dB, 0 dB, 2 dB, and 6 dB) using pure heart sound as the target signal and pure lung heart sound as the noise signal. All the sounds were converted into spectral-domain by applying the short-time Fourier transform (STFT) with a 2048 frame length and 128 frame shifts. Because high frequency parts may not provide critical information for further analyses, we only use 0-300 bins (corresponding to 0-1170 Hz) in this study.

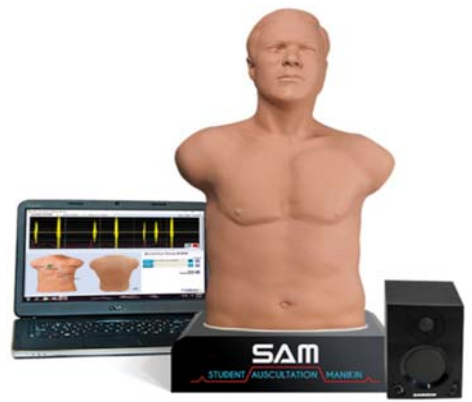

Fig. 5. Student Auscultation Manikin (SAM).

*B. Latent space analysis of a selected case*

In this section, we used a sample mixed sound to detail every step in the PC-DAE system. Fig. 6 shows the overall procedure of the PC-DAE, where Fig. 6(a) and (b) show the spectrograms of pure heart and lung sounds, respectively. Fig. 6 (c) shows the latent representation extraction process. For demonstration purpose, we selected two specific neurons, one corresponding to heart sounds and the other corresponding to lung sounds, and plotted their trajectories along the time axis in Fig. 6(d) and (e), respectively. By evaluating Fig. 6(d) and (e), we first perceive that the periodicity properties of Fig. 6(d) and (e) aligned well with Fig. 6(a) and (b), respectively. Meanwhile, we observe different trajectories of these two neurons, and the periodicity of heart sound is different from lung sound. Next, we applied the DFT on the trajectories of Fig. 6(d) and (e) and obtained Fig. 6 (f) and (g), respectively, to capture the periodicity more explicitly. Notably, the x-axis for Fig. 6(a), (b), (d), and (e) is time (s), while the x-axis of Fig. 6(f) and (g) is frequency (Hz). In the temporal signal analysis, the signals in Fig. 6(f) and (g) are termed MFA [59] of Fig. 6 (d) and (e). As can be seen by converting the trajectory into the modulation domain, the periodicity can be more easily observed.

(0-1); (f) and (g) is the DFT results, where the x-axis is the frequency and y-axis denotes the power density.

By comparing Fig. 6(f) and (g), we observe a peak in the low-frequency part in Fig. 6(g), and a peak is located at a high-frequency part in Fig. 6(f). The results suggest that these two neurons should be clustered into two different groups. We apply the same procedures (trajectory extraction and DFT) on all the neurons in the DAE. The neurons that process shorter and longer periodicity are clustered into two distinct groups. Finally, given a mixed sound, we first extract the latent representation; to extract heart sounds, we then keep the neurons that correspond to heart sounds and deactivated the neuron that corresponds to lung sounds and vice versa.

To further verify the effectiveness of the PC clustering approach, we compare DC and PC clustering approaches by qualitatively analyzing the clustering results. To facilitate a clear visual comparison, we adopted the principle component analysis (PCA) [60] to reduce the dimensions on the latent representations to only 2-D and then draw the scattering plots in Fig. 7. The figure shows the spectrograms of two mixed heart-lungs sounds and the clustering results of latent representations. Fig. 7(a) shows the spectrogram of a mixed normal heart sound and abnormal lung (rhonchi) sound; Fig. 7(b) shows the spectrogram of a mixed normal heart sound and abnormal lung (stridor) sound. Fig. 7(c) and (d) are the DC clustering results of latent representations (dimensionality-reduced by PCA) corresponding to Fig. 7(a) and (b), respectively. Fig. 7(e) and (f) are the PC clustering results of the latent representations corresponding to Fig. 7(a) and (b), respectively.

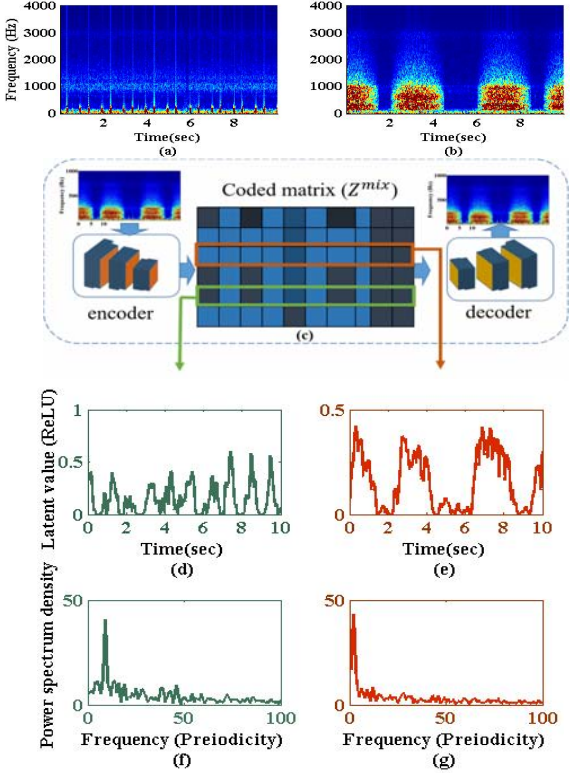

Fig. 6. Analyses of latent representations of sample sounds. (a) and (b), respectively, are the spectrograms of the pure heart and lung sounds, the x-axis is time (s) and y-axis is frequency (Hz); (c) presents the latent representation extraction based on the DAE model; (d) and (e) are trajectories of two latent neurons, where the x-axis is the time, and the y-axis is activation value

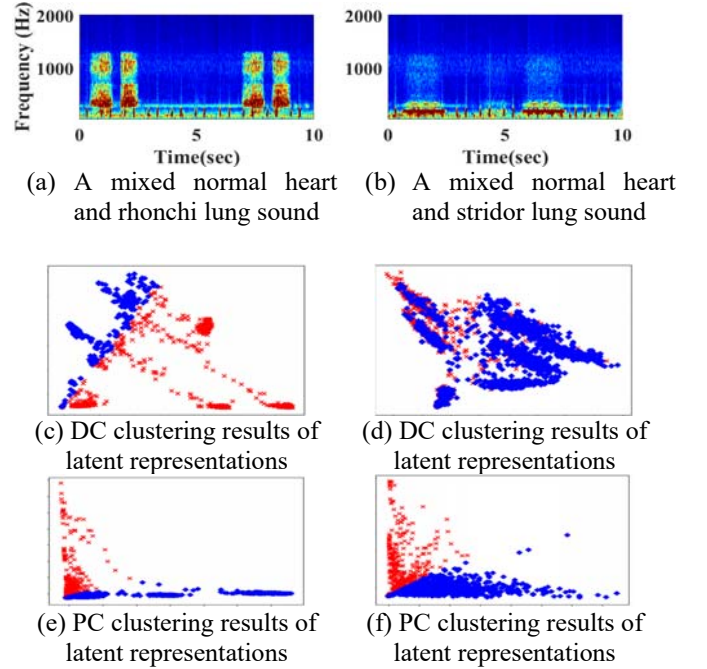

(a) A mixed normal heart and rhonchi lung sound
(b) A mixed normal heart and stridor lung sound
(c) DC clustering results of latent representations
(d) DC clustering results of latent representations
(e) PC clustering results of latent representations
(f) PC clustering results of latent representations

Fig. 7. Spectrograms of two mixed heart-lung sounds and the clustering results of latent representations. (a) and (b) are the spectrograms of two mixed heart and lung sounds; (c) and (d) are the DC clustering results of the latent representation; (e) and (f) are the PC clustering results of the latent representation.

Table 1

Evaluation results of separated heart sounds generated by the proposed PC-DAE(F) and PC-DAE(C) comparing to three conventional approaches in terms of SDR, SIR, and SAR. Avg denotes the average scores over five SNRs.

|      | DC-NMF | | | PC-NMF | | | DC-DAE(C) | | | PC-DAE(F) | | | PC-DAE(C) | | |
|------|-----|-----|-----|-----|-----|-----|-----|-----|-----|-----|-----|-----|-----|-----|-----|
|      | SDR | SIR | SAR | SDR | SIR | SAR | SDR | SIR | SAR | SDR | SIR | SAR | SDR | SIR | SAR |
| -6dB | -2.74 | 0.22 | 5.19 | -2.08 | 1.09 | 6.09 | 0.86 | 2.02 | 11.24 | -1.01 | 0.32 | 9.58 | 2.37 | 3.57 | 12.00 |
| -2dB | -0.45 | 3.76 | 4.54 | -0.92 | 2.66 | 6.60 | 3.01 | 4.58 | 10.91 | 2.36 | 3.94 | 10.51 | 6.46 | 7.81 | 13.78 |
| 0dB  | 0.29 | 5.04 | 4.61 | -1.12 | 2.17 | 5.87 | 5.68 | 7.60 | 13.31 | 3.52 | 5.96 | 10.43 | 7.57 | 9.17 | 14.32 |
| 2dB  | 0.81 | 5.71 | 4.69 | 1.75 | 6.14 | 6.97 | 6.35 | 7.97 | 14.23 | 5.59 | 8.14 | 10.89 | 9.38 | 10.97 | 15.40 |
| 6dB  | 2.49 | 9.09 | 4.70 | 4.82 | 11.19 | 7.88 | 8.49 | 11.07 | 14.04 | 7.54 | 11.70 | 10.46 | 12.54 | 14.97 | 16.79 |
| Avg  | 0.08 | 4.76 | 4.75 | 0.49 | 4.65 | 6.68 | 4.88 | 6.65 | 12.75 | 3.60 | 6.01 | 10.38 | 8.72 | 10.44 | 14.95 |

Table 2

Evaluation results of separated lung sounds generated by the proposed PC-DAE(F) and PC-DAE(C) comparing to three conventional approaches in terms of SDR, SIR, and SAR. Avg denotes the average scores over five SNRs.

|      | DC-NMF | | | PC-NMF | | | DC-DAE(C) | | | PC-DAE(F) | | | PC-DAE(C) | | |
|------|-----|-----|-----|-----|-----|-----|-----|-----|-----|-----|-----|-----|-----|-----|-----|
|      | SDR | SIR | SAR | SDR | SIR | SAR | SDR | SIR | SAR | SDR | SIR | SAR | SDR | SIR | SAR |
| -6dB | -2.71 | -0.01 | 5.28 | -2.48 | 0.54 | 5.94 | -0.97 | 0.23 | 10.68 | -1.15 | 0.11 | 9.39 | 3.40 | 4.64 | 12.69 |
| -2dB | -0.02 | 3.62 | 5.78 | -0.52 | 3.32 | 5.80 | 3.11 | 5.10 | 11.37 | 2.70 | 4.26 | 10.73 | 6.94 | 9.45 | 13.14 |
| 0dB  | 1.04 | 4.80 | 6.00 | 0.93 | 6.25 | 5.45 | 4.82 | 6.70 | 13.12 | 3.40 | 5.10 | 11.25 | 8.17 | 10.82 | 14.11 |
| 2dB  | 2.44 | 6.89 | 6.07 | 1.84 | 7.58 | 5.40 | 5.62 | 8.50 | 11.66 | 5.56 | 7.86 | 11.65 | 9.16 | 12.05 | 13.96 |
| 6dB  | 3.34 | 9.21 | 6.08 | 3.34 | 9.66 | 6.26 | 8.50 | 12.02 | 12.45 | 8.30 | 11.61 | 12.10 | 10.90 | 14.88 | 14.16 |
| Avg  | 0.82 | 4.90 | 5.84 | 0.62 | 5.47 | 5.77 | 4.22 | 6.51 | 11.86 | 3.76 | 5.79 | 11.02 | 7.71 | 10.37 | 13.61 |

By observing Fig. 7(a), (c), and (e), we can note that heart and lung sounds showed clearly different time-frequency properties (as shown in Fig. 7(a)). In this case, both DC (as in Fig. 7(c)) and PC (as shown in Fig. 7(e)) clustering approaches can effectively group the latent features corresponding to lung and heart sounds into two distinct groups. Consequently, satisfactory separation results can be achieved for both DC and PC approaches. Next, by observing the results of Fig. 7(b), (d), and (f), since the stridor sound are highly overlapped with heart sound (as show in Fig. 7(b)), the DC clustering approach (as show in Fig. 7(d)) cannot effectively group the latent representations into two distinct groups. On the other hand, the PC clustering approach (as show in Fig. 7(f)) can successfully cluster the latent representations into two distinct groups and consequently yield better separation results.

Please note that any particular time-frequency representation method can be used to perform MFA. The present study adopts the DFT as a representative method. Other time-frequency representation methods, such as CWT [29-31][61] and Hilbert–Huang transform [62-64], can be used. When using these methods, suitable basis functions or prior knowledge need to be carefully considered. In this study, we intend to focus our attention on DFT and will further explore other time-frequency representation methods in the future.

*C. Quantitative evaluation based on source separation evaluation metrics*

Next, we intend to compare the separation performance using Eqs. (9) and (10) and Eqs. (13) and (14). The results are listed in Fig. 8. Since Eqs. (9) and (10) directly estimate the hear sound and lung sounds, the results using Eqs. (9) and (10) are termed "Direct". On the other hand, because Eqs. (13) and (14) estimate the heart and lung sounds by a ratio mask function, results are termed "Mask". We tested the performance using both PC-DAE(F) and PC-DAE(C). From the results in Fig. 8, we observe the results of "Mask" consistently outperform that of "Direct" except for heart sound's SIR of PC-DAE(F), and confirm the effectiveness of using a ratio mask function to perform separation instead of direct estimation. In the following discussion, we only report the PC-DAE separation results using the ratio mask functions of Eqs. (13) and (14).

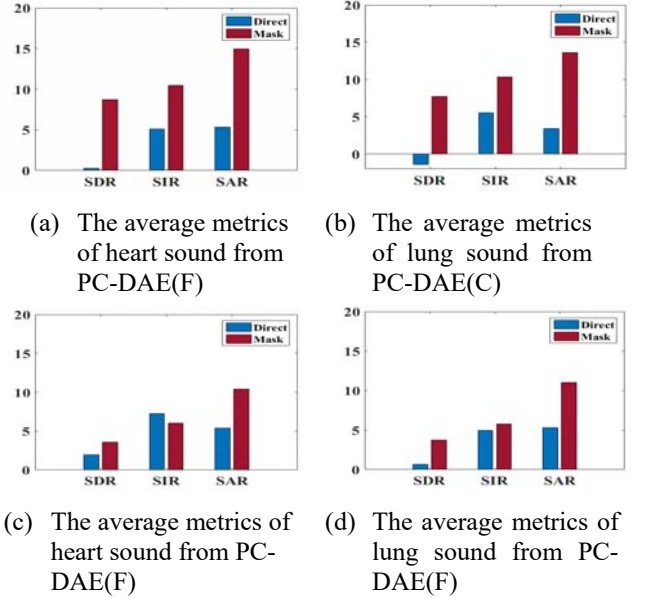

(a) The average metrics of heart sound from PC-DAE(F)

(b) The average metrics of lung sound from PC-DAE(C)

(c) The average metrics of heart sound from PC-DAE(F)

(d) The average metrics of lung sound from PC-DAE(F)

Fig. 8. Average separation results over different SNR conditions. (a) and (c) show the heart sound separation results using PC-DAE(C) and PC-DAE(F), respectively; (b) and (d) show the lung sound separation results using PC-DAE(C) and PC-DAE(F), respectively.

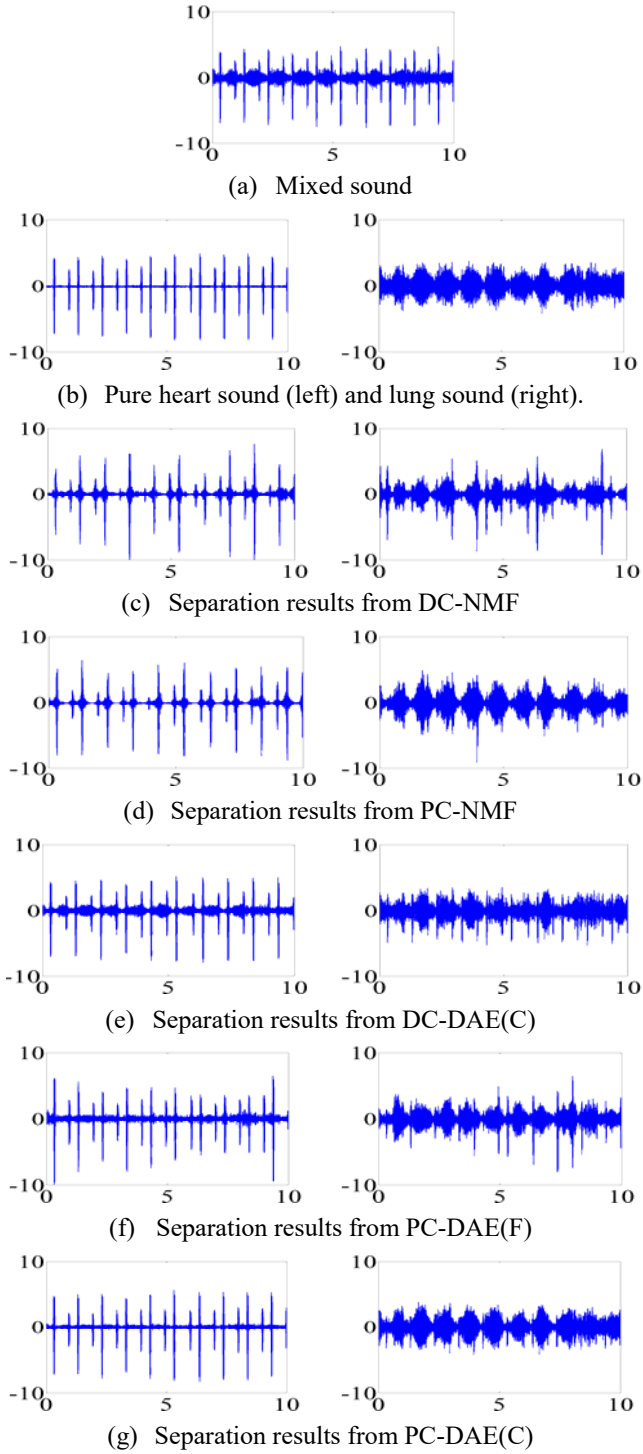
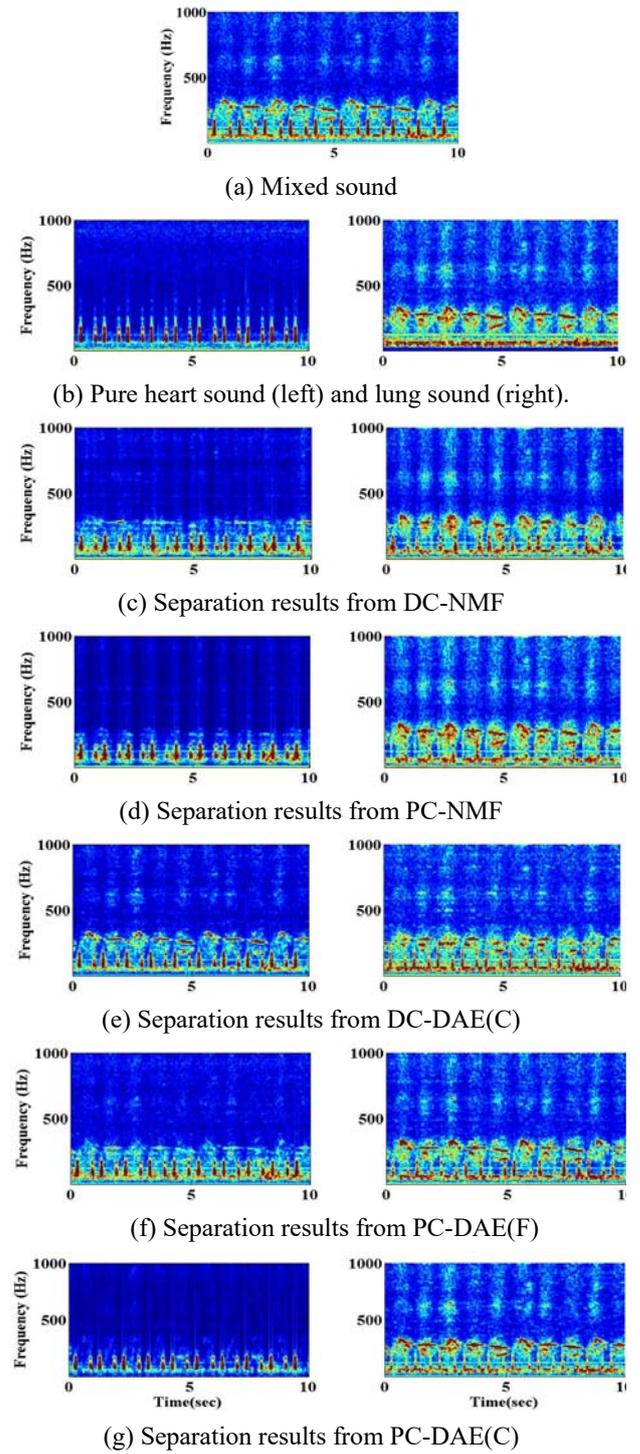

Fig. 9. The waveform of a mixed sample. The y-axis is the amplitude of the signals, and the x-axis is time index (s). From (b) to (g), the left and right panels are heart sound and lung sound, respectively.

Fig. 10. Spectrograms of a mixed sample. The y-axis is the frequency of the signals, and the x-axis is time index (s). From (b) to (g), the left and right panels are heart sound and lung sound, respectively.

Tables 1 and 2 show the evaluation results of heart and lung sounds, respectively, tested on the proposed PC-DAE(F) and PC-DAE(C) with comparative methods. The separation performance is consistent for heart and lung sounds. From the two tables, we observe all the SDR, SIR, and SAR scores mostly increase along with increasing SNR levels. Meanwhile, we note that PC-NMF outperforms DC-NMF, and PC-DAE(C) outperforms DC-DAE(C), confirming the periodicity property to provide superior separation performance than direct clustering. Meanwhile, we observed that the deep learning-based approaches, namely DC-DAE(C) and PC-DAE(C), outperform NMF-based counterparts, namely DC-NMF and PC-NMF, verifying the effectiveness of deep learning models to extract representative features over shallow models. Finally, we observe that PC-DAE(C) outperforms PC-DAE(F), suggesting that the convolutional architecture can yield superior performance than fully connected architecture for this sound separation task.

*D. Qualitative comparison based on separated waveforms and spectrograms*

In addition to quantitative comparison, we also demonstrate waveforms and spectrums of a sample sound to visually compare the separation results. We selected a sample sound, which is the mixed sound with the SNR ratio of heart sound (treated as the signal) and wheezing lung sound (treated as the noise) to be 6 dB. Fig. 9 demonstrates the waveforms of the sample sound, where Fig. 9(a) shows the mixed sounds. Fig. 9(b) shows the pure heart sound (left panel) and lung sound (right panel) that have not been mixed. Fig. 9(c), (d), (e), (f), and (g) show the separated results of DC-NMF, PC-NMF, DC-DAE(C), PC-DAE(F), and PC-DAE(C), respectively. From Fig. 9, we observe that PC-DAE(C) can more effectively separate the heart and lung sounds as compared to other methods; the trends are consistent with those shown in Tables 1 and 2.

Next in Fig. 10, we show the spectrograms of the same sample sound shown in Fig. 9. Fig. 10(a) presents the mixed sounds, Fig. 10(b) shows the pure heart and lung sounds, and Fig. 10(c) to (g) are separated results. From Fig. 10(a), we can observe that the two sounds are highly overlapped in the lower frequency region. It is also noticed that PC-NMF possesses a higher performance for interference suppression during the high frequency of lung sounds, and PC-DAE(F) possesses a higher performance in overlapped frequency bandwidth and receives improved heart sound quality. PC-DAE(F) and PC-DAE(C) performed the best with minimal artificial noises. Generally speaking, the two PC-DAE approaches outperformed the other approaches yielding clear separation spectrograms.

Table 3
Recognition accuracies of mixed heart-lung sounds and separated heart sounds with different age and gender groups.

|  | Mixed heart-lung sound | | | | | | Separated heart sound | | | | | |
| --- | --- | --- | --- | --- | --- | --- | --- | --- | --- | --- | --- | --- |
| Age | 0-20 | | 21-65 | | 66-80 | | 0-20 | | 21-65 | | 66-80 | |
| Gender | Male | Female | Male | Female | Male | Female | Male | Female | Male | Female | Male | Female |
| Accuracy | 71% | 67% | 67% | 76% | 71% | 57% | 81% | 90% | 80% | 85% | 86% | 90% |
| Avg | 69% | | 72% | | 64% | | 86% | | 83% | | 88% | |

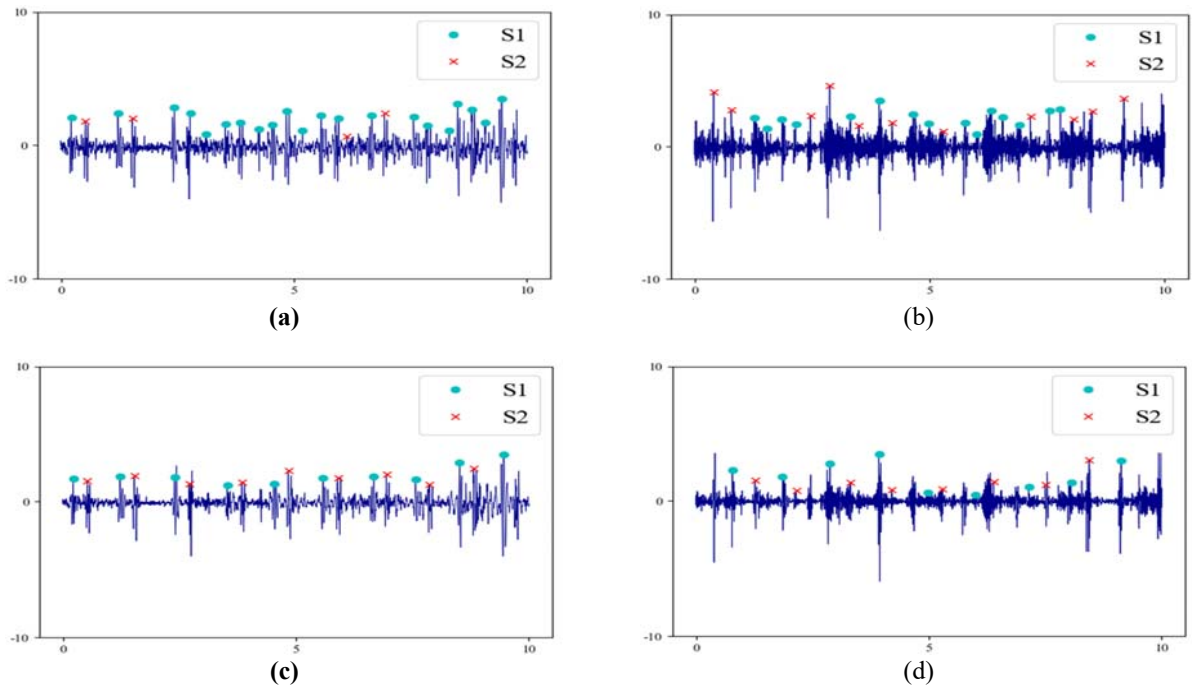

Fig 11. The waveforms of two sound samples and the corresponding S1-S2 recognition results. (a) a mixed heart-lung sound with normal heart sound and normal lung sound. (b) a mixed heart-lung sound with abnormal heart sound and abnormal lung sound. (c) and (d) are the separated results corresponding to (a) and (b), resepctively. The recognized S1 and S2 results are colored by green and red symbols, respectively.

*E. Real application in first heart sound (S1) and second heart sound (S2) recognition*

We used another dataset to further evaluate the proposed algorithm in a more real-world scenario. Real mixed heart-lung sounds were collected from National Taiwan Hospital, and the proposed PC-DAE was used to separate the heart and lung sounds. Because it is not possible to access pure heart and lung sounds corresponding to the mixed heart-lung sounds, the SDR, SIR, and SAR scores cannot be used as the evaluation metrics in this task. Instead, we adopted the first heart sound (S1) and second heart sound(S2) recognition metric accuracies to determine the separation performance. We adopted a well-known S1 and S2 recognition algorithm from [10, 65], which considers frequency properties and the assumption of S1-S2 and S2-S1 intervals. We believe that this alternative metric is convincing and valuable since the S1-S2 recognition accuracy has already been used as a crucial index for doctors to diagnose the occurrence of diseases[66, 67].

This dataset includes 3 different age groups, namely 0-20 (childhood and adolescence), 21-65 (adulthood), and over 66 (senior citizen)). Each group has 6 cases, including 3 males and 3 females, and each case has 7 mixed heart-lung sounds (10 sec). Based on this design, we can determine whether the proposed approach can be robust against variations of age and gender groups (accordingly covering people with different physiological factors, such as blood pressure, heart rate, etc.). Table. 3 shows the recognition accuracies of before and after performing heart-lung sound separation.

To visually investigate the S1-S2 recognition performance, we present the waveforms along with the recognition results in Fig. 11. Fig 11 (a) and (b) are two sound samples, where Fig. 11 (a) is the mixed heart-lung sound with normal heart and lung sounds, and Fig. 11 (b) is the mixed heart-lung sound with abnormal heart sound (weak periodicity) and abnormal lung sound (rhonchi). Fig 11 (c) and (d) show the S1-S2 recognition the after performing heart-lung sound separation corresponding to Fig 11 (a) and (b), respectively.

From Fig. 11 (a) and (b), we can note that the S1-S2 recognition results are poor for the mixed sounds, and the recognition performance are notably improved with the separated heart sounds (as can be seen from Fig 11 (c) and (d)), confirming the effect of the proposed PC-DAE's outstanding capability of separating the heart sounds from mixed sounds.

## V. CONCLUSION

The proposed PC-DAE is derived based on the periodicity properties of the signal to perform blind source separation in a single-channel recording scenario. Different from the conventional supervised source separation approach, PC-DAE does not require supervised training data. To the best of our knowledge, the proposed PC-DAE is the first work that combines the advantages of deep-learning-based feature representations and the periodicity property to carry out heart-lung sound separations. The results of this study indicate that the proposed method is effective to use a periodic analysis algorithm to improve the separation of sounds with overlapped frequency bandwidth. The results also show that PC-DAE provided satisfactory separation results and achieve superior quality as compared to several related works. Moreover, we verified that by using the proposed PC-DAE as a preprocessing step, the heart sound recognition accuracies can be considerably improved. In our current work, we need to define how many sources are in the signal. However, in most cases, determining the exact number of the sources is difficult. Hence, identifying on effective way to determine the number of the sources is an important future work. In the present study, we consider the condition where only sounds recorded by an electronic stethoscope is available. We believe that this experiment setup is close to most real-world clinical scenarios. In the future, we will extend the proposed PC-DAE to the conditions where additional physiological data is available, such as ECG, photoplethysmogram, and blood pressure signals.